\documentclass{article}
\usepackage[utf8]{inputenc}
\usepackage{amsmath}
\usepackage{bm}
\title{An implicit constitutive relation in which the stress and the linearized strain appear linearly, for describing the small displacement gradient response of elastic solids}
\author{K. R. Rajagopal\\
Department of Mechancial Engineering\\
Texas A\&M Univeristy\\
College Station, Texas-77845}

\usepackage[round]{natbib}

\usepackage{color,soul}

\date{December 2020}

\begin{document}

\maketitle \
 
\begin{abstract}
In this short note we develop a constitutive relation that is linear in both the Cauchy stress and the linearized strain, by linearizing implicit constitutive relations between the stress and the deformation gradient that have been put into place to describe the response of elastic bodies (see \cite{rajagopal2003implicit}), by assuming that the displacement gradient is small. These implicit equations include the classical linearized elastic constitutive approximation as well as constitutive relations that imply limiting strain, as special subclasses. 
\end{abstract}
\section{Introduction}
The classical linearized theory of elasticity is derived as an approximation within the context of the Cauchy theory of elasticity (see \cite{cauchy1822recherches}, \cite{cauchy1828equations}) wherein the Cauchy stress $\mathbf{T}$ is expressed explicitly as a function of the deformation gradient $\mathbf{F}$, under the assumption that the norm of the displacement gradient is sufficiently small. This linearization that stems from the Cauchy theory of elasticity, while astonishingly successful in describing the response of many elastic solids when they are sustaining small displacement gradients, has several shortcomings: (a) The theory is not in keeping with the demands of causality in that it is the body and surface forces and consequently the stress in the body that cause the body to deform rather than the deformation causing the stresses, and thus it would be reasonable to expect the deformation gradient or the strain to be specified in terms of the stress and not vice-versa. Interestingly the approximate constitutive relation describing linearized elastic response is expressed by defining the stress in terms of the linearized strain and also the linearized strain in terms of the stress. However, the representation for the linearized strain in terms of the stress does not stem from an appropriate approximation of a nonlinear theory wherein a proper measure of the deformation is assumed to depend nonlinearly on the stress. (b) Many metallic alloys (see \cite{saito2003multifunctional}, \cite{sakaguch2004tensile}, \cite{sakaguchi2005effect},\cite{hao2005super}, \cite{li2007ideal}, \cite{talling2008determination}, \cite{withey2008deformation}, \cite{zhang2009fatigue}) as well as ubiquitous and commonly used materials like concrete (see \cite{grasley2015model}) exhibit nonlinear response even in the range of “strains” that are considered small wherein classical linearized elasticity is supposed to be operative. The classical linearized theory is  incapable and impotent to describe the nonlinear response that is observed in the aforementioned materials. (c) The material moduli that go into characterizing linearized elastic response, the Lame’ moduli (or equivalently the Young’s modulus, and Poisson’s ratio) have to be constants (they cannot depend on the density, strain, mean value of the stress, etc.) as the theory would not then be linear in the stress and the linearized strain (the density in virtue of the balance of mass depends on the linearized strain). Often one sees ad hoc assumptions concerning properties such as the Young's modulus depending on the pressure(the mean value of the stress) but this is not possible within the context of the classical linearized theory of elasticity. (d) Since the stress and linearized strain are related linearly, if the stress becomes very large as in problems such as the state of stress at a crack tip or due to a concentrated load (namely loading that leads to singularity in the stresses), the strains have to necessarily become very large, thereby contradicting the basic tenet under which the linearization is established. (e) The linearized elastic response cannot capture the possibility of limited small strain as the stress increases. The consequences of above drawbacks are discussed in detail in \cite{rajagopal2014nonlinear}.The above comments concerning the linearization that stems from Cauchy elasticity are also true of the linearization that stems from Green elasticity (see \cite{green1837laws}, \cite{green1841propagation}) as Green elasticity is a sub-class of Cauchy elasticity wherein one assumes the existence of a stored energy function that depends on the deformation gradient.\\

The question which then confronts us is whether we can develop a constitutive relation for elastic bodies which when linearized under the assumption that the displacement gradients are small does not present the shortcomings discussed above, and the answer to that question is a resounding yes. Recently, Rajagopal (see \cite{rajagopal2003implicit}, \cite{rajagopal2007elasticity}, \cite{rajagopal2011conspectus}) showed that the class of bodies that are incapable of dissipation in the sense that they cannot convert mechanical working into heat (energy in thermal form) is much larger than the class of Cauchy elastic bodies, and \cite{rajagopal2007response} provided a thermodynamic basis for the development of implicit constitutive relations for such bodies. These implicit constitutive relations include amongst other subclasses, constitutive relations for classical Cauchy elastic bodies as well constitutive relations wherein the deformation gradient is expressed as a function of the stress. It transpires that linearization of the the latter class of constitutive relations leads to an approximation wherein the linearized elastic strain is a nonlinear function of the Cauchy stress (see \cite{devendiran2017thermodynamically}, \cite{sandeep2016numerical} and \cite{kulvait2017modeling} for experimental corroboration of such approximations of the nonlinear constitutive relations). Included amongst this class are constitutive relations wherein the Cauchy stress and  the linearized strain appear linearly.\\

The equation that will be considered in this short note are not bilinear pairings or bilinear functions as usually defined (see \cite{MacLane1967Algebra}) as this requires the pairing $ f(x,y)$ to be both left linear in $x$ and right linear in $y$, since the constitutive relations that we consider also include linear terms including bilinear terms. \\

2. Implicit constitutive relations in which both the stress and linearized strain appear linearly. \\

The starting point for the study of elastic bodies described by implicit constitutive relations is (see \cite{rajagopal2003implicit})\\
\begin{eqnarray}
\label{Implicit Equation 1}
\mathbf{f(\rho,T,F,\bm{X})=0}
\end{eqnarray}
where $\rho$ is the density, $ \bm{T}$ is the Cauchy stress, $ \bm{F}$ is the deformation gradient and $\bm{X}$ is a point in the reference configuration.  This point just acts as a surrogate for a particle belonging to the real body. The body is inhomogeneous if the response is different at different points belonging to the real body and it is homogeneous if the response is the same at every particle of the body((see \cite{noll1958mathematical} and \cite{truesdell2004non} for a detailed and careful discussion of material isomorphism, material uniformity, and homogeneity)). Henceforth, we shall not indicate the dependence of the function of $\bm{X}$ for the sake of notational simplicity.\\

In the case of isotropic elastic bodies, the above implicit relation reduces to
 \begin{eqnarray}
 \label{Implicit Equation 2}
\mathbf{f(\rho, T,B)=0},
\end{eqnarray}         
where $\mathbf{B}$ is the right Cauchy-Green tensor. One can use representation theorems and obtain the relationship between the stress and the Cauchy-Green tensor $\mathbf{B}$. In the case of the constitutive relation (2), standard representation theory leads to (see \cite{spencer1975theory})
 \begin{multline}
 \alpha_0\bm{I}+\alpha_1\bm{T}+\alpha_2\bm{B}+\alpha_3\bm{B^2}+\alpha_4\bm{T^2}+\alpha_5(\bm{TB+BT})+\alpha_6(\bm{TB^2+B^2T})\\
 +\alpha_7(\bm{T^2B+BT^2})+\alpha_8(\bm{B^2T^2+T^2B^2)}=\bm{0}.
\end{multline}
where $\alpha_i, i=1,----8,$ depend on
$\rho, tr\bm{T}, tr\bm{B}, tr\bm{T^2}, tr\bm{B^2}, tr\bm{(T^3)}, tr\bm{B^3},\\ tr\bm{TB}, tr\bm{T^2B}, tr\bm{B^2T}, tr\bm{T^2B^2}.$\\

Implicit constitutive relations developed for describing the elastic response of bodies, when linearized, allow the material moduli that describe the body to depend on the density, the invariants of the stress, invariants of the linearized strain as well as mixed invariants (see \cite{rajagopal2003implicit}, \cite{rajagopal2007elasticity}), they however have to depend on the density and the invariants of the linearized strain in a very special way.\\ 

Let us carry out the linearization by assuming that\\
 \begin{eqnarray}
 \max_{\bm{X}\in B, t\in R} \left\|\dfrac{\partial\bm{u}}{\partial x}\right\|=O(\delta), \qquad \delta<<1.
\end{eqnarray}

A special sub-class of such elastic bodies in given by
 \begin{eqnarray}
 \label{Sub-class of Elastic Bodies}
 \bm\beta_0{\epsilon}+\beta_1 \mathbf{I}+\beta_2 \bm{T}+\beta_3\mathbf{T^2}+\beta_4[
\bm{T \epsilon}+\bm{\epsilon T}]+\beta_5[\bm{T^2\epsilon+\epsilon T^2}]=\bm{0}
 \end{eqnarray}
 where the $\beta_i, i=1,2,3$ are scalar valued functions that can at most depend linearly on $\bm{\epsilon}$, but arbitrarily on the invariants of $\bm{T}$, while $\beta_i, i=0,4,5$ depend on the invariants of $\bm{T}$. Because the density depends on the trace of epsilon we cannot have the above constants depend in a random fashion on the density if we want the model to depend linearly on $\bm{\epsilon}$.\\
 
 We notice that equation (5) in general provides an implicit nonlinear relationship between the linearized strain and Cauchy stress. A special subclass of such models is given by constitutive relations that provide an explicit nonlinear expression for the linearized strain as a function of the Cauchy stress. It is possible that some of these nonlinear relations can be inverted to obtain the stress as a nonlinear function of the linearized strain. Such an expression does not have the status of an approximate constitutive relation, as discussed later, it is a purely mathematical expression that can be manipulated, but ultimately it has to be expressed as the linearized strain as a function of stress (see \cite{rajagopal2018note} for a detailed discussion of linearization of implicit constitutive relations). In fact, it is best not to invert the expression for $\epsilon$ as a function of $\sigma$ even if it were possible as it might mislead one in the misapplication and misinterpretation of the procedure and results.\\
 
 If we require that the implicit constitutive relation in which both the stress and the linearized strain appear linearly, then (5) reduces to
 \begin{multline}
   E_1\bm{\epsilon}+E_2 \bm{T}+E_3(tr\bm{\epsilon} ){\bm{T}+E_4 (tr\bm{T})\bm{I}}+E_5(tr\bm{\epsilon)}\bm{I}+E_6 (tr\bm{T})\bm{\epsilon}\\
+ E_7 (tr\bm{T})(tr \bm{\epsilon})\bm{I}+E_8(\bm{\epsilon} \bm{T}+\bm{T}\bm{\epsilon})+E_9(tr(\bm{\epsilon}\bm{T})\bm{I})=\bm{0}.   
 \end{multline}
 
In the above equation, $ E_i, i=1--8 $ are constants. Let us suppose that $E_1$ is not zero and divide through by $E_1$ to express the above equation as
\begin{multline}
\bm{\epsilon}+A_1 \bm{T}+A_2(tr\bm{\epsilon} ){\bm{T}+A_3 (tr\bm{T})\bm{I}}+A_4(tr\bm{\epsilon)}\bm{I}+A_5 (tr\bm{T})\bm{\epsilon}\\
+A_6 (tr\bm{T})(tr \bm{\epsilon})\bm{I}+A_7(\bm{\epsilon} \bm{T}+\bm{T}\bm{\epsilon})+A_8tr(\bm{\epsilon}\bm{T})\bm{I}=\bm{0}.    
\end{multline}
In the above equation $ A_i, i=1--7$ are constants.\\

While the scalar valued functions can be viewed as material moduli that characterize the body described by the implicit constitutive relation, one could also view the constitutive relation (5) from a different perspective. Since the balance of mass
\begin{eqnarray}
\rho_R=\rho(det\bm{F}), 
\end{eqnarray}
where $\rho_R$ is the density in the reference configuration and $\rho$ the density in the deformed current configuration, can be approximated in the case of small displacement gradients , within the context of (4), as 
\begin{eqnarray}
\rho_R=\rho(1+tr\bm{\epsilon}),
\end{eqnarray}
we can replace $tr\bm{\epsilon} $ by the density, That is, in virtue of the balance of mass, we  can think of $tr\bm{\epsilon}$ as a substitute for the density $\rho$, and thus one can view, for instance, $E_2 (tr\bm{\epsilon})$ as a density dependent material moduli. Of course, since $tr\bm{\epsilon}$ can only occur linearly, the dependence of the material moduli on the density has to be in a special manner. An interesting consequence of recognizing that material moduli depending linearly on the $tr\bm{\epsilon}$ allows one to consider the response of inhomogeneous bodies wherein the material moduli can depend on the density, which in turn can depend on the particle in the reference configuration. Using such an approach, \cite{murru2020density}, \cite{murru2020density2} have recently studied damage in concrete.\\

When $A_2, A_4, A_5, A_6, A_7$ and $A_8$ are zero, the model reduces to
\begin{eqnarray}
\bm{\epsilon}=-{A_1}{\bm{T}-{A_3}(tr\bm{T})\bm{I}}, 
\end{eqnarray}
and we can then identify $A_1$ and $A_3$ as
\begin{eqnarray}
{A_1}=-\frac{(1+\nu)}{E} ; {A_3}=\frac{\nu}{E},
\end{eqnarray}
 where $E$ is the Young's modulus and $\nu$ is the Poisson's ratio.\\

3. A Strain Limiting Constitutive Relation.\\

In order to illustrate some interesting features of the constitutive relation (7), let us consider the one dimensional static response wherein the stress and the linearized strain take the form:
\begin{eqnarray}
\bm{T}=\sigma(\bm{e_1\otimes\bm{e_1}}),\bm{\epsilon}=\epsilon(\bm{e_1\otimes\bm{e_1})},
\end{eqnarray}
where $\bm{e_i}$, i=1,2,3 denote unit vectors in a Cartesian co-ordinate system.\\

On substituting (12) into (7) and simplifying, we obtain
\begin{eqnarray}
\epsilon=\frac{-(A_1+A_3)\sigma}{[(1+A_4)+(A_2+A_5+A_6+2A_7+A_8)\sigma]}.
\end{eqnarray}
First, let us suppose that $ (A_2+A_5+A_6+2A_7+A_8)$ is not zero. Then, when $ \sigma=0$, we find $\epsilon=0$ and when $\epsilon=0$, we find that $\sigma=0 $, which is to be expected. Moreover, we note that when $ \sigma $ tends to infinity, we notice that
\begin{eqnarray}
\epsilon=\frac{-[A_1+A_3]}{(A_2+A_5+A_6+2A_7+A_8)},
\end{eqnarray}
that is there is a limit to the strain. We need this limit to be positive. Since $ A_1+A_3=-1/E$, which is negative, we need the denominator $ (A_2+A_3+A_6+2A_7+A_8)$ to be positive. Next, we note that when the stress is compressive, if $ \sigma=-\frac{(1+A_4)}{ (A_2+A_3+A_6+2A_7+A_8)} $ then the linearized strain blows up, but this is not allowed. Hence we need to ensure that the compressive stress is always greater than $-\frac{(1+A_4)}{ (A_2+A_3+A_6+2A_7+A_8)}$. One could view limiting the range of stresses that are allowable as a tremendous drawback invalidating the use of such a constitutive relation. To the contrary, the constitutive relation that we are dealing with is an approximation and the constitutive relation is valid only if we ensure that the basic approximation that we have made is in place, namely that the strains have to be sufficiently small. Thus, depending on the values of the material parameters only a certain range of values for the compressive stresses would be allowable. When $ (A_2+A_5+A_6+2A_7+A_8)=0$, the constitutive relation reduces to the approximate classical linearized elastic constitutive relation and as we know such a constitutive relation does not exhibit limited strain (Of course, the linearized constitutive relation ought not to be used when the strains are larger than the requirement expressed by (4). However,ignoring such a caveat the linearized constitutive relation is used to study problems wherein singularities in stresses, which implies singularities in the linearized strains, arise).\\

It is important to recognize that one can also obtain an expression for the stress $ \sigma $ in terms of $ \epsilon $ by manipulating equation (14), but one ought to recognize that this does not lead to an appropriate approximate constitutive relation. In general an expression for the Cauchy-Green stretch as a nonlinear function of the Cauchy stress when inverted and then linearized will not lead to the same constitutive relation that is obtained by linearizing and then inverting. The process of linearization just with respect to the displacement gradient and inversion are not commutative operations when dealing with these implicit equations. That is, an expression for the Cauchy Green tensor $ \bm{B} $ as a nonlinear function of the Cauchy stress $ \bm{T} $, when linearized and then inverted will not give rise to the same model as the process of inverting and then linearizing, when the linearization is carried out just with respect to the displacement gradient. Thus, even if one can obtain a nonlinear expression for the Cauchy stress $ \bm{T} $ in terms of $ \bm{\epsilon} $ it can never arise from a linearization of the Cauchy stress within the context of the Cauchy theory of elasticity as on linearization within the Cauchy theory of elasticity, one is inexorably and inescapably led to the linearized approximate constitutive equation. Thus, if one is going to use the linearization of implicit equations wherein one obtains a nonlinear expression for the linearized strain in terms of the stress, it is best not to invert it and obtain the expression for the stress in terms of the linearized strain. However, if one does that, one has to then remember that such an expression is valid only for strains wherein the original expression is valid. \\

 Let us now turn our attention to a special sub-class of constitutive relations of (7) that take the form:
 \begin{eqnarray}
(1+\lambda_1(tr\bm{T}))\bm{\epsilon}=B_1(1+\lambda_2(tr\bm{\epsilon}))\bm{T}+B_2(1+\lambda_3(tr\bm{\epsilon)})(tr\bm{T})\bm{I}.
 \end{eqnarray}
 When $\lambda_i, i=1,2,3$ are zero and when\\
 \begin{eqnarray}
 {B_1}=\frac{(1+\nu)}{E},{B_2}=\frac{-\nu}{E},
 \end{eqnarray}
 where E is the Young's modulus and $\nu$ is the Poisson's ratio, we recover the classical linearized elastic model.\\
 
 Also, when $ \lambda_1 $ is zero in (15), the constitutive relation reduces to
 \begin{eqnarray}
\bm{\epsilon}=B_1(1+\lambda_2(tr\bm{\epsilon}))\bm{T}+B_2(1+\lambda_3(tr\bm{\epsilon}))(tr\bm{T})\bm{I}.
 \end{eqnarray}

 Next, we can express the constitutive relation (15) in a manner in which the dependence on density is brought out explicitly. In virtue of (9), and (4) it follows that
 \begin{eqnarray}
 \rho= \rho_R(1-tr\bm{\epsilon}),
 \end{eqnarray}
and thus
 \begin{eqnarray}
 (1+\lambda_2tr\bm{\epsilon})=1+\frac{\lambda_2(\rho_R-\rho)}{\rho_R}\\
 (1+\lambda_3tr\bm{\epsilon})=1+\frac{\lambda_3(\rho_R-\rho)}{\rho_R}.
 \end{eqnarray}
 Hence, equation (17) can be rewritten as
 \begin{eqnarray}
 \bm{\epsilon}=B_1[(1+\lambda_2)-(\lambda_2)\frac{\rho}{\rho_R}]\bm{T}+B_2[(1+\lambda_3)-\lambda_3\frac{\rho}{\rho_R})](tr\bm{T})\bm{I}.
 \end{eqnarray}
 The above simple constitutive equation in which the stress, the linearized strain and the density occur linearly accords an opportunity to study several interesting and important classes of problems.\\
 
 First, let us consider the problem of damage which is essentially a consequence of the inhomogeneity of the body when the material properties deteriorate as the body is deformed. The constitutive relation (22) provides a simple method to incorporate the deterioration of material properties in a body and the resulting degradation and damage that ensues. When $ \lambda_2$ and $\lambda_3$ are negative, we notice that the product of $B_1$ and $B_2$ with the term adjacent to them in the square parenthesis can be viewed as material parameters that decrease as the density decreases. Of course, depending on whether $\lambda_2$ and $\lambda_3$ are positive or negative and the deformation in question and thus the density, locally material properties may increase or decrease in value. Using a similar model, Murru et al. (2020a), (2020b) studied damage that takes place in cement concrete. Their results are in qualitative agreement with experimental observation. This is just one of many problems wherein damage occurs even within small strains, in inhomogeneous bodies, that can be studied within the context of such simple models in which the stress, linearized strain and density occur linearly by assuming that properties vary with $\bm{X}$.\\

Finally, it is worth considering unsteady deformations of a body described by the constitutive relation (6), especially with regard to the possibility of unsteady motions. For the sake of illustration, let us consider one-dimensional problems.\\

First, let us record the equations that govern the motion of a body described by the one-dimensional constitutive relation corresponding to (15):
\begin{eqnarray}
(1+\lambda_1 \sigma)\epsilon=B_1(1+\lambda_2\epsilon)\sigma+B_2(1+\lambda_3\epsilon)\sigma,
\end{eqnarray}
where we have assumed that the stress $ \bm{T}$ and $\bm{\epsilon}$ are of the form given by (12), but now under the assumption that
\begin{eqnarray}
\sigma=\sigma (x,t), \epsilon=\epsilon (x,t).
\end{eqnarray}
The balance of linear momentum in one dimensions, in the absence of body forces, reduces to
\begin{eqnarray}
\rho(\dfrac{\partial^2u}{\partial t^2})=\dfrac{\partial\sigma}{\partial  x}.
\end{eqnarray}
Next, let us define
\begin{eqnarray}
\alpha=(B_1\lambda_2+B_2\lambda_3).
\end{eqnarray}
Then, equation (17) can be expressed as 
\begin{eqnarray}
\epsilon=\frac{\sigma}{E}+\alpha\epsilon\sigma.
\end{eqnarray}
However, one could also express (17) as
\begin{eqnarray}
\sigma=\frac{E\epsilon}{(1+E\alpha\epsilon)}=E\epsilon (1-E\alpha\epsilon),
\end{eqnarray}
the last equality is a consequence of (4) which implies $ \epsilon $ is small provided we assume that $ E\alpha $ is of order 1. As we shall see later, using the above expression for the stress $\sigma $\ in terms of $\epsilon$ needs to be interpreted carefully though it accords some ease with manipulating the equations as it will be seem below.
We notice that the stress $ \sigma$ is a nonlinear function of $ \epsilon$  and this is not tenable as an appropriate constitutive relation even as an approximation. As mentioned earlier, we can use the expression for mathematical manipulations but we then have to re-invert to interpret the results within the context of the approximation (15).\\

Now, on inserting the expression for $ \sigma $ given in (27) into the right hand side of the equation of motion, we obtain 
\begin{eqnarray}
\rho(\dfrac{\partial^2u}{\partial t^2})=E\dfrac{\partial}{\partial x}[\epsilon(1-E\alpha\epsilon)]
\end{eqnarray}
and since
\begin{eqnarray}
\epsilon=\dfrac{\partial u}{\partial x},
\end{eqnarray}
we obtain
\begin{eqnarray}
\rho(\dfrac{\partial^2u}{\partial t^2})=E\dfrac{\partial }{\partial x}[\dfrac{\partial u}{\partial x}(1-E\alpha\dfrac{\partial u}{\partial x})]
\end{eqnarray}
which can be expressed as $
\rho(\dfrac{\partial^2u}{\partial t^2})=E\dfrac{\partial^2 u}{\partial x^2}-E^2\alpha\dfrac{\partial}{\partial x}[(\dfrac{\partial u}{\partial x})^2]
$\\
and when $ \alpha $ is zero we recover the one dimensional wave equation. We need to supply the initial and boundary conditions in order to solve a specific problem of interest.\\
Instead of choosing to express the stress $\sigma$ as a function of the linearized strain $\epsilon$ the ideal way to address the problem is to solve equation (24) and the following simultaneously:
\begin{eqnarray}
\frac{\partial u}{\partial x}=\frac{\sigma}{E(1-\alpha\sigma)}.
\end{eqnarray}
Let us now consider the possibility of unsteady motion in a one-dimensional body described by the constitutive relation (6). Let us once again assume that the stress and strain are given by (12) with $ \sigma$ and $ \epsilon$ given by (23). In this case we obtain an expression for the stress $ \sigma$ in terms of $ \epsilon$ and substitute the same into the balance of linear momentum. However, as mentioned before, it is better to not invert the expression for the linearized strain in terms of the stress but express the stress as a funtion of the linearized strain (also, it might not always be possible to invert the expression for the linearized strain as a function of the stress). \\
We now find that 
\begin{multline}
\sigma=-\frac{(1+A_4)\epsilon}{[(A_1+A_3)+(A_2+A_5+A_6+2A_7+A_8)\epsilon]}=\frac{E\epsilon(1+A_4)}{[1+E\beta\epsilon]}\\=E\epsilon(1+A_4)(1-E\beta\epsilon) 
\end{multline}
where we have used $ -(A_1+A_3)= \frac{1}{E}$, that $ \epsilon $ is small and $ \beta$ defined through
\begin{eqnarray}
\beta=\frac{A_2+A_5+A_6+2A_7+A_8}{E}
\end{eqnarray}
and $ E\beta$ is of order 1. Substituting (32) into (24),and using (29) we obtain\\
\begin{eqnarray}
\rho(\dfrac{\partial^2u}{\partial^2t})=E\dfrac{\partial}{\partial x}[(1+A_4)\frac{\partial u}{\partial x}(1-E\beta\frac{\partial u}{\partial x})]=E(1+A_4)[\dfrac{\partial^2 u}{\partial x^2}-E\beta\dfrac{\partial}{\partial x}[(\dfrac{\partial u}{\partial x})^2].
\end{eqnarray}\\

Once again, it is better to solve (24) simultaneously with
\begin{eqnarray}
\frac{\partial u}{\partial x}=\frac{-(A_1+A_3)\sigma}{(1+A_4)+(A_@+A_5+A_6+2A_7+A_8)\sigma}
\end{eqnarray}
simultaneously.\\

Acknowledgement\\
K. R. Rajagopal thanks the Office of Naval Research for support of this work.\

\bibliographystyle{plainnat}

\begin{thebibliography}{28}
\providecommand{\natexlab}[1]{#1}
\providecommand{\url}[1]{\texttt{#1}}
\expandafter\ifx\csname urlstyle\endcsname\relax
  \providecommand{\doi}[1]{doi: #1}\else
  \providecommand{\doi}{doi: \begingroup \urlstyle{rm}\Url}\fi

\bibitem[Cauchy(1823)]{cauchy1822recherches}
AL~Cauchy.
\newblock Recherches sur l'{\'e}quilibre et le mouvement int{\'e}rieur des
  corps solides ou fluides, {\'e}lastiques ou non {\'e}lastiques.
\newblock \emph{Bull. Soc. Philomath}, 9-13=Oeuvres (2), 2, 1823.

\bibitem[Cauchy(1828)]{cauchy1828equations}
AL~Cauchy.
\newblock \emph{Sur les {\'e}quations qui expriment les conditions
  d’{\'e}quilibre ou les lois du mouvement int{\'e}rieur d’un corps solide,
  {\'e}lastique, ou non {\'e}lastique}, volume~3.
\newblock 1828.

\bibitem[Devendiran et~al.(2017)Devendiran, Sandeep, Kannan, and
  Rajagopal]{devendiran2017thermodynamically}
VK~Devendiran, RK~Sandeep, K~Kannan, and KR~Rajagopal.
\newblock A thermodynamically consistent constitutive equation for describing
  the response exhibited by several alloys and the study of a meaningful
  physical problem.
\newblock \emph{International Journal of Solids and Structures}, 108:\penalty0
  1--10, 2017.

\bibitem[Grasley et~al.(2015)Grasley, El-Helou, D’Ambrosia, Mokarem, Moen,
  and Rajagopal]{grasley2015model}
Z~Grasley, R~El-Helou, M~D’Ambrosia, D~Mokarem, C~Moen, and KR~Rajagopal.
\newblock Model of infinitesimal nonlinear elastic response of concrete
  subjected to uniaxial compression.
\newblock \emph{Journal of Engineering Mechanics}, 141:\penalty0 04015008,
  2015.

\bibitem[Green(1837)]{green1837laws}
G~Green.
\newblock On the laws of the reflexion and refraction of light at the common
  surface of two non-crystallized media.
\newblock \emph{Trans Cambr Phil Soc 1839}, 7(1839–1842): 1-24 Mathematical
  papers of the late George Green, edited by N. M. Ferris, 245–269 MacMillan
  and Company, London (1871)., 1837.

\bibitem[Green(1839)]{green1841propagation}
G~Green.
\newblock On the propagation of light in crystallized media.
\newblock \emph{Trans Cambr Phil Soc}, 7 (1839-1842), 121-140= Papers 293-311
  (1841), 1839.

\bibitem[Hao et~al.(2005)Hao, Li, Sun, Zheng, Hu, and Yang]{hao2005super}
YL~Hao, SJ~Li, SY~Sun, CY~Zheng, QM~Hu, and R~Yang.
\newblock Super-elastic titanium alloy with unstable plastic deformation.
\newblock \emph{Applied Physics Letters}, 87\penalty0 (9):\penalty0
  091906--1--091906--3, 2005.

\bibitem[Kulvait et~al.(2017)Kulvait, M{\'a}lek, and
  Rajagopal]{kulvait2017modeling}
V~Kulvait, J~M{\'a}lek, and KR~Rajagopal.
\newblock Modeling gum metal and other newly developed titanium alloys within a
  new class of constitutive relations for elastic bodies.
\newblock \emph{Archives of Mechanics}, 69:\penalty0 223--241, 2017.

\bibitem[Li et~al.(2007)Li, Morris~Jr, Nagasako, Kuramoto, and
  Chrzan]{li2007ideal}
T~Li, JW~Morris~Jr, N~Nagasako, S~Kuramoto, and DC~Chrzan.
\newblock “ideal” engineering alloys.
\newblock \emph{Physical review letters}, 98:\penalty0 105503, 2007.

\bibitem[MacLane and Birkhoff(1967)]{MacLane1967Algebra}
S~MacLane and G~Birkhoff.
\newblock \emph{Algebra}.
\newblock The MacMillan Company, Collier MacMillan, London, 1967.

\bibitem[Murru et~al.(2020{\natexlab{a}})Murru, Torrence, Grasley, Rajagopal,
  Alagappan, and Garboczi]{murru2020density}
PT~Murru, C~Torrence, Z~Grasley, KR~Rajagopal, P~Alagappan, and E~Garboczi.
\newblock Density-driven damage mechanics (d3-m) model for concrete i:
  mechanical damage.
\newblock \emph{International Journal of Pavement Engineering}, pages 1--14,
  2020{\natexlab{a}}.

\bibitem[Murru et~al.(2020{\natexlab{b}})Murru, Torrence, Grasley, Rajagopal,
  Alagappan, and Garboczi]{murru2020density2}
PT~Murru, C~Torrence, Z~Grasley, KR~Rajagopal, P~Alagappan, and E~Garboczi.
\newblock Density-driven damage mechanics (d3-m) model for concrete ii:
  mechanical damage.
\newblock \emph{International Journal of Pavement Engineering},
  2020{\natexlab{b}}.

\bibitem[Noll(1958)]{noll1958mathematical}
W~Noll.
\newblock A mathematical theory of the mechanical behavior of continuous media.
\newblock \emph{Archive for rational Mechanics and Analysis},
  197--226,2,Reprinted in Rational Mechanics of Materials, Intl. Sci. Rev.
  Series, Gordon and Breach, New York (1965), 1958.

\bibitem[Rajagopal(2003)]{rajagopal2003implicit}
KR~Rajagopal.
\newblock On implicit constitutive theories.
\newblock \emph{Applications of Mathematics}, 28:\penalty0 279--319, 2003.

\bibitem[Rajagopal(2007)]{rajagopal2007elasticity}
KR~Rajagopal.
\newblock The elasticity of elasticity.
\newblock \emph{Zeitschrift f{\"u}r angewandte Mathematik und Physik},
  58\penalty0 (2):\penalty0 309--317, 2007.

\bibitem[Rajagopal(2011)]{rajagopal2011conspectus}
KR~Rajagopal.
\newblock Conspectus of concepts of elasticity.
\newblock \emph{Mathematics and Mechanics of Solids}, 16:\penalty0 536--562,
  2011.

\bibitem[Rajagopal(2014)]{rajagopal2014nonlinear}
KR~Rajagopal.
\newblock On the nonlinear elastic response of bodies in the small strain
  range.
\newblock \emph{Acta Mechanica}, 225:\penalty0 1545--1553, 2014.

\bibitem[Rajagopal(2018)]{rajagopal2018note}
KR~Rajagopal.
\newblock A note on the linearization of the constitutive relations of
  non-linear elastic bodies.
\newblock \emph{Mechanics Research Communications}, 93:\penalty0 132--137,
  2018.

\bibitem[Rajagopal and Srinivasa(2007)]{rajagopal2007response}
KR~Rajagopal and AR~Srinivasa.
\newblock On the response of non-dissipative solids.
\newblock \emph{Proceedings of the Royal Society A: Mathematical, Physical and
  Engineering Sciences}, 463\penalty0 (2078):\penalty0 357--367, 2007.

\bibitem[Saito et~al.(2003)Saito, Furuta, Hwang, Kuramoto, Nishino, Suzuki,
  Chen, Yamada, Ito, Seno, Nonaka, Ikehata, Nagasako, Iwamoto, Ikuhara, and
  Sakuma]{saito2003multifunctional}
T~Saito, T~Furuta, JH~Hwang, S~Kuramoto, K~Nishino, N~Suzuki, R~Chen, A~Yamada,
  K~Ito, Y~Seno, T~Nonaka, H~Ikehata, N~Nagasako, C~Iwamoto, Y~Ikuhara, and
  T~Sakuma.
\newblock Multifunctional alloys obtained via a dislocation-free plastic
  deformation mechanism.
\newblock \emph{Science}, 300:\penalty0 464--467, 2003.

\bibitem[Sakaguch et~al.(2004)Sakaguch, Niinomi, and
  Akahori]{sakaguch2004tensile}
N~Sakaguch, M~Niinomi, and T~Akahori.
\newblock Tensile deformation behavior of ti-nb-ta-zr biomedical alloys.
\newblock \emph{Materials transactions}, 45:\penalty0 1113--1119, 2004.

\bibitem[Sakaguchi et~al.(2005)Sakaguchi, Niinomi, Akahori, Takeda, and
  Toda]{sakaguchi2005effect}
N~Sakaguchi, M~Niinomi, T~Akahori, J~Takeda, and H~Toda.
\newblock Effect of ta content on mechanical properties of ti--30nb--xta--5zr.
\newblock \emph{Materials Science and Engineering: C}, 25:\penalty0 370--376,
  2005.

\bibitem[Sandeep et~al.(2016)Sandeep, Kannan, and
  Rajagopal]{sandeep2016numerical}
RK~Sandeep, K~Kannan, and KR~Rajagopal.
\newblock Numerical and approximate analytical solutions for cylindrical and
  spherical annuli for a new class of elastic materials.
\newblock \emph{Archive of Applied Mechanics}, 86:\penalty0 1815--1826, 2016.

\bibitem[Spencer(1975)]{spencer1975theory}
AJM Spencer.
\newblock Theory of invariants,[in:] continuum physics, eringen ac.
\newblock \emph{Academic Press, New York}, 2, 1975.

\bibitem[Talling et~al.(2008)Talling, Dashwood, Jackson, Kuramoto, and
  Dye]{talling2008determination}
RJ~Talling, RJ~Dashwood, M~Jackson, S~Kuramoto, and D~Dye.
\newblock Determination of (c11-c12) in ti--36nb--2ta--3zr--0.3 o (wt.\%)(gum
  metal).
\newblock \emph{Scripta Materialia}, 59\penalty0 (6):\penalty0 669--672, 2008.

\bibitem[Truesdell and Noll(1992)]{truesdell2004non}
C~Truesdell and W~Noll.
\newblock The non-linear field theories of mechanics.
\newblock In \emph{The non-linear field theories of mechanics}.
  Springer-Verlag, Berlin-Heidelber-New York, 1992.

\bibitem[Withey et~al.(2008)Withey, Jin, Minor, Kuramoto, Chrzan, and
  Morris~Jr]{withey2008deformation}
E~Withey, M~Jin, A~Minor, S~Kuramoto, DC~Chrzan, and JW~Morris~Jr.
\newblock The deformation of “gum metal” in nanoindentation.
\newblock \emph{Materials Science and Engineering: A}, 493:\penalty0 26--32,
  2008.

\bibitem[Zhang et~al.(2009)Zhang, Li, Jia, Hao, and Yang]{zhang2009fatigue}
SQ~Zhang, SJ~Li, MT~Jia, YL~Hao, and R~Yang.
\newblock Fatigue properties of a multifunctional titanium alloy exhibiting
  nonlinear elastic deformation behavior.
\newblock \emph{Scripta Materialia}, 60\penalty0 (8):\penalty0 733--736, 2009.

\end{thebibliography}

\end{document}